# A scalar arterial input function does not uniquely determine the perfusion residue: Theory and phantom MRI

Yuyang Gan[1,2], Dominick Romano[1,3], Maneesh John[1,4], Renjiu Hu[1,2], Qihao Zhang[1],

Benjamin Weppner[1,3], Alexey V. Dimov[1], Pascal Spincemaille[1], Yi Wang[1,2,3,4,*]

1 Department of Radiology, Weill Cornell Medicine, New York, NY

2 Mechanical and Aerospace Engineering, Cornell University, Ithaca, NY

3 Biomedical Engineering, Cornell University, Ithaca, NY

4 Electrical and Computer Engineering, Cornell University, Ithaca, NY

* Correspondence to: Yi Wang, PhD, Department of Radiology, Weill Cornell Medicine, 407 East 61st Street, RR-118, New York, NY 10065, USA.

E-mail: yiwang@med.cornell.edu

**Running head:** Scalar AIF Does Not Uniquely Determine Residue

**Word count:** 231 (Abstract, excluding subheadings; 237 including subheadings), 2798 (Body, Introduction through Conclusions)

## Abstract

### Purpose

To relate the residue function of tracer kinetic models to blood velocity and cross-sectional tracer distributions at the arterial inlet, and show that a cross-sectionally averaged arterial input function (AIF) does not uniquely determine the residue.

### Theory and Methods

For steady laminar flow without radial diffusion, the residue function referenced to the cross-sectionally averaged inlet AIF was analytically derived from the velocity and inlet tracer distributions. In a straight tube with parabolic flow, two nonnegative inlet profiles sharing one measured AIF were compared for gamma-variate and rectangular arterial spin labeling (ASL) inputs. A straight-tube phantom was imaged at 3 T during gadolinium bolus passage, with uniform fills as controls.

### Results

In the phantom, enhancement during bolus arrival was greater near the lumen center than near the wall at all 11 stations and for all three boluses, whereas uniform gadolinium fills showed no such pattern. In the tube model, the two profiles produced tissue curves differing by up to 23.3% of the uniform-inlet peak and 0.30 s in time to peak for the gamma-variate input, and by 33.3% for ASL labels shorter than the fastest transit time. Their residues differed in initial amplitude and normalized temporal shape.

### Conclusion

Identical AIFs can produce different tissue concentration curves, depending on the spatial distributions of inlet tracer concentration and arterial flow velocity. Consequently, the residue-AIF tracer kinetic models may not be consistent with fluid mechanics or interpreted as tissue flow.

## Introduction

Tissue blood flow, important for assessing many diseases, is encoded in dynamic tracer imaging, including dynamic contrast CT, dynamic susceptibility contrast (DSC) MRI, dynamic contrast-enhanced (DCE) MRI (1), and arterial spin labeling (ASL) (2). Extraction or quantification of tissue blood flow has been based on the tracer kinetic model or indicator-dilution theory: the tissue concentration-time curve is modeled as blood flow multiplied by the temporal convolution of an arterial input function (AIF) (3,4) with a tissue residue function (5), and blood flow is estimated by temporal deconvolution. Multiple compartment models express a balance of tracer entry and exit using the same kinetics based on the AIF (6). However, the AIF reflects only a single concentration-time curve and does not resolve the tracer spatial distribution across the arterial lumen.

Physically, the velocity field of blood is governed by the Navier-Stokes equations, and the transport of tracer from the feeding artery to the tissue by the advection-diffusion (continuity) equation (7). Classical indicator-dilution theory proposes that tissue concentration is the temporal convolution of the residue with the AIF (8), whereas imaging measures a cross-sectionally averaged AIF that does not record how tracer is distributed across streamlines of different velocities. Because tracer entering near the axis leaves sooner than tracer entering near the wall, the tissue response depends on this inlet distribution. Here we show that, even for a fixed vascular geometry and velocity field, a cross-sectionally averaged AIF does not uniquely determine the residue function. In general, a residue as a temporal function may not be derivable from fluid mechanics in complex tissue vasculature. We prove this analytically for a straight tube and image radial tracer gradients during bolus passage in a flow phantom.

## Theory and Methods

### Theory: Arterial input and tissue residue function

The standard intravascular indicator-dilution model relates the tissue tracer concentration $c_T(t)$ to an AIF, $c_A(t)$, through blood flow per unit tissue volume $(f)$ and a dimensionless residue function $r(\tau)$, with $r(0) = 1$ (8):

$$c_T(t) = f \int_0^t c_A(t-\tau)\, r(\tau)\, \mathrm{d}\tau. \tag{1}$$

The temporal integral sums all earlier arterial inputs still in the tissue, each weighted by the residue function $r$ (4,8); for an exponential residue function, $r(\tau) = e^{-k\tau}$, with rate constant $k$ (6), Eq. 1 is equivalent to the compartment model $\frac{\mathrm{d}}{\mathrm{d}t} c_T(t) = f \cdot c_A(t) - k \cdot c_T(t)$. Supporting Information S1 derives Eq. 1 from tracer mass conservation.

In this work, we demonstrate that a given AIF can lead to different residue functions $r$ when the inlet tracer distributions are changed, even when vascular geometry and velocity fields are fixed. To do so, we consider a simple vascular space, a cylindrical tube (9), as an all-vascular model region: $c_A(t)$ is the concentration averaged over the inlet cross-section and $c_T(t)$ the concentration averaged over the volume of the tube. $c_T(t)$ plays the role of the tissue concentration in Eq. 1.

Let $R$ be the radius of this tube and $L$ its length. Let $c(z, \rho, \theta, t)$ be the tracer concentration at axial position $0 \le z \le L$, radial position $0 \le \rho \le R$, and azimuthal angle $0 \le \theta < 2\pi$. Let $S = \pi R^2$ be its cross-sectional area and $V = SL$ its volume. Flow is steady and axial with velocity $u(\rho)$; the volumetric flow is $F = \int_S u(\rho)\, \mathrm{d}S$, so $f = F/V$. Radial diffusion is neglected, so each tracer element remains on its streamline, as in convective laminar-flow dilution theory (10). If the tube initially contains no tracer and the inlet concentration is zero for $t < 0$, transport through length $L$ simply delays the inlet concentration on each streamline by its own travel time (11):

$$c(z,\rho,\theta,t) = c\left(0,\rho,\theta,t - \frac{z}{u(\rho)}\right). \tag{2}$$

The measured AIF is the area mean over the inlet surface, and the tissue concentration is the volume mean over the tube:

$$c_A(t) = \frac{1}{S}\int_S c(0,\rho,\theta,t)\,\mathrm{d}S,$$

$$c_T(t) = \frac{1}{V}\int_V c(z,\rho,\theta,t)\,\mathrm{d}V, \tag{3}$$

with $\mathrm{d}S = \rho\,\mathrm{d}\rho\,\mathrm{d}\theta$ and $\mathrm{d}V = \rho\,\mathrm{d}\rho\,\mathrm{d}\theta\,\mathrm{d}z$ in cylindrical coordinates. Consider an inlet distribution whose cross-sectional shape does not change during the bolus: $c(0,\rho,\theta,t) = \phi(\rho,\theta)c_A(t)$, with $\frac{1}{S}\int_S \phi(\rho,\theta)\,\mathrm{d}S = 1$, so that $c_A$ remains the area-mean inlet concentration. The derivation holds for any profile $\phi(\rho,\theta)$, axially symmetric or not. Substituting into Eq. 3 gives

$$\begin{aligned}
c_T(t) &= \frac{1}{V}\int_V \phi(\rho,\theta)c_A\left(t - \frac{z}{u(\rho)}\right)\,dV \\
&= \frac{1}{V}\int_V \phi(\rho,\theta)\int_{-\infty}^{\infty}\delta\left(\tau - \frac{z}{u(\rho)}\right)c_A(t-\tau)\mathrm{d}\tau\,\mathrm{d}V \\
&= \int_{-\infty}^{\infty} c_A(t-\tau)\int_V \frac{1}{V}\phi(\rho,\theta)\,\delta\left(\tau - \frac{z}{u(\rho)}\right)\mathrm{d}V\,\mathrm{d}\tau\,,
\end{aligned}$$

where $\delta$ is the delta function. The quantity $\frac{z}{u(\rho)}$ is the time a tracer element at $(z,\rho,\theta)$ has spent in the tube since entering it. Because we set $c_A(t) = 0$ for $t < 0$, this integral becomes

$$c_T(t) = f\int_0^t c_A(t-\tau)\,\tilde{r}_\phi(\tau)\,\mathrm{d}\tau. \tag{4}$$

Here

$$\tilde{r}_\phi(\tau) = \frac{1}{F}\int_0^L\int_0^{2\pi}\int_0^R \phi(\rho,\theta)\,\delta\left(\tau - \frac{z}{u(\rho)}\right)\rho\,\mathrm{d}\rho\,\mathrm{d}\theta\,\mathrm{d}z = \frac{1}{F}\int_0^{2\pi}\int_0^R \phi(\rho,\theta)\,u(\rho)\,H\left(\frac{L}{u(\rho)} - \tau\right)\rho\,\mathrm{d}\rho\,\mathrm{d}\theta\,, \tag{5}$$

with $H$ the Heaviside function (for $\tau > 0$). The quantity $\tilde{r}_\phi$ is an effective residue referenced to the measured AIF $c_A$ and depends on the velocity distribution $u$ and the inlet shape $\phi$.

Integrating along each streamline, $\int_0^L \delta\left(\tau - \frac{z}{u(\rho)}\right) \mathrm{d}z = u(\rho) H\left(\frac{L}{u(\rho)} - \tau\right)$ for $\tau > 0$, gives the second form, where $\phi u\, \rho\, \mathrm{d}\rho\, \mathrm{d}\theta$ is the tracer flux entering through the area element $\rho\, \mathrm{d}\rho\, \mathrm{d}\theta$ per unit $c_A$, and the Heaviside function $H$ keeps only streamlines whose transit time $\frac{L}{u(\rho)}$ exceeds $\tau$, that is, streamlines on which tracer that entered a time $\tau$ earlier is still inside the tube. For $\tau$ shorter than the fastest transit time, all streamlines contribute, and Eq. 5 gives $\tilde{r}_\phi(\tau) = \kappa_\phi \equiv \frac{1}{F} \int_0^{2\pi} \int_0^R \phi u\, \rho\, \mathrm{d}\rho\, \mathrm{d}\theta$, the ratio of the flux-weighted to the area-weighted inlet concentration. This ratio equals 1 for a uniform inlet profile (Supporting Information S1). Dividing by it gives the normalized residue

$$R_\phi(\tau) = \frac{\tilde{r}_\phi(\tau)}{\kappa_\phi} = \frac{\int_0^{2\pi} \int_0^R \phi u\, H\left(\frac{L}{u} - \tau\right) \rho\, \mathrm{d}\rho\, \mathrm{d}\theta}{\int_0^{2\pi} \int_0^R \phi u\, \rho\, \mathrm{d}\rho\, \mathrm{d}\theta}, \tag{6}$$

which is the fraction of entered tracer that remains in the tube a time $\tau$ after entry, with $R_\phi(0) = 1$ as for the standard residue function of Eq. 1. Eq. 4 is thus Eq. 1 with the normalized residue $R_\phi$ and the flux-weighted inlet concentration $\kappa_\phi c_A$ replacing $r$ and $c_A$, respectively. The dependence on $\phi$ is not limited to the scale factor $\kappa_\phi$: the shape of $R_\phi$ also changes with $\phi$. Integrating Eq. 5 over $\tau$ gives $\int_0^\infty \tilde{r}_\phi(\tau)\, \mathrm{d}\tau = \frac{L}{F} \int_0^{2\pi} \int_0^R \phi\, \rho\, \mathrm{d}\rho\, \mathrm{d}\theta = \frac{V}{F}$ for every inlet profile with unit area mean; the area under the effective residue is set by the vascular volume and flow alone. Two profiles with different $\kappa_\phi$, and therefore different initial heights, cannot have proportional effective residues; their normalized residues (integrating to $V/(\kappa_\phi F)$) must differ in shape. Each fixed inlet profile thus has its own scalar convolution (Eq. 4), but no single residue function can describe all possible inlet profiles that share the same measured AIF.

**Methods: Phantom and MRI acquisition**

A phantom was constructed using a straight polyvinyl chloride tube (nominal inner diameter 9.53 mm, length 81.3 cm). Through this tube, water flowed driven by gravity at approximately 200 mL/min (Re $\approx$ 450). Its lumen, larger than that of cerebral arteries, allowed cross-sectional imaging with a clinical dynamic sequence. Three boluses of gadopentetate dimeglumine (Magnevist; Bayer Corporation, Whippany, NJ), each nominally 2 mL at a prepared concentration of 5 mM, were injected manually through a Y-connector over approximately 5 s. The main analysis used the first bolus, the only one entering a tracer-free baseline; all three are compared in Supporting Information S2.

Imaging used a 3-T GE Discovery MR750 scanner (GE HealthCare, Waukesha, WI) and a 3D DISCO/EFGRE3D sequence: TR/TE 4.22/1.636 ms, flip angle 10°, in-plane acceleration factor 2, reconstructed matrix 256 × 256 × 20, and voxel size 1.0156 × 1.0156 × 1.0 $mm^3$. Each series contained 121 volumes reconstructed at 1.530-s intervals; because of view sharing, the effective temporal footprint of each volume is longer than this interval. Five acquisitions used the same protocol: static water, flowing water, bolus injection, static diluted gadolinium, and flowing diluted gadolinium. Because gadolinium filled the tube uniformly in the last two, any radial pattern in them reflects imaging effects, such as flip-angle nonuniformity or flow effects on the signal, rather than bolus transport.

Cross-sections perpendicular to the tube centerline were resampled on a 0.4-mm grid. Elliptical lumen boundaries were fitted independently to the static- and flowing-water acquisitions (median center difference 0.12 mm; median Dice overlap 0.98) and averaged into a single lumen geometry, which was fixed before any gadolinium images were sampled and used for all comparisons. Eleven stations with fully imaged cross-sections were analyzed; clipped sections were excluded (Supporting Information S3). Figure 1A–B show a 3.0-mm longitudinal slab along this centerline and the water-derived lumen mask.

Because the signal was not calibrated, tracer concentration as a function of position $\boldsymbol{r}$ and time $t$ was represented by the relative enhancement of the MR signal $S(\boldsymbol{r}, t)$:

$$E(\boldsymbol{r}, t) = \frac{S(\boldsymbol{r},t) - S_0(\boldsymbol{r})}{S_0(\boldsymbol{r})}. \tag{7}$$

Bolus maps used the pre-bolus median (frames 4–8, zero-based) of the injection scan as baseline to avoid inter-scan drift. Quantitative comparisons referenced the bolus and both uniform fills to one common baseline (mean of flowing-water frames 20–99), averaging uniform fills over those frames; each station's first bolus was evaluated on its local upslope (30% of lumen-mean peak to peak, up to six frames).

At each station, we computed ten-bin radial enhancement profiles, the difference between mean enhancement in a central band (labeled "core" in the figures, $\rho/R < 0.30$) and an inner wall band ($0.60 \leq \rho/R < 0.85$), and the distance of the positive-enhancement centroid from the lumen center. Here $\rho/R$ is the radial position normalized to the fitted elliptical lumen boundary (0 at the center, 1 at the wall); the wall band stops short of the boundary to avoid partial-volume voxels, and the centroid is weighted by positive $E$. Station values from this single experiment are reported descriptively, without significance tests.

## Results

### Phantom: nonuniform enhancement across the lumen

Bolus arrival produced greater enhancement near the center of the tube than near its wall (15.3 s; Figure 1A,B, top). During washout the pattern reversed, with enhancement concentrated near the wall (39.8 s; Figure 1B, bottom), where tracer lingered on slow near-wall streamlines. This evolution is expected for laminar transport and shows that the cross-sectional profile of a real bolus changes during its passage. The tube model below takes the simplest case, a fixed inlet profile, for which a scalar convolution still holds but the residue is not unique.

Referenced to the common flowing-water baseline, the bolus radial profile declined from 0.38 near the axis to 0.15 at the lumen edge, whereas the static and flowing gadolinium fills coincided near 0.45 without central enhancement (Figure 1C). Central-minus-wall enhancement averaged 0.108 for the bolus and was positive at all 11 stations, versus −0.019 and −0.027 for the static and flowing fills (Figure 1D). The positive-enhancement centroid was displaced by 1.26 mm for the bolus versus 0.78 and 0.74 mm for the fills (Figure 1E). The larger bolus offset is consistent with injection through the off-axis Y-connector port. Because the fills' central-minus-wall values were slightly negative, imaging effects did not create the central excess. Cross-sectional averaging, as in a measured AIF, erases these spatial variations.

### Tube model: identical AIFs, different residue functions

Assuming steady parabolic (Poiseuille) flow

$$u(\rho) = u_{\max}[1 - (\frac{\rho}{R})^2],$$

we get $F = S u_{\max}/2$.

Defining $s = (\rho/R)^2$ and $\tau_0 = L/u_{\max}$, the transit time along the axis (the fastest streamline), and writing $\phi(s)$ for an axially symmetric inlet profile expressed in $s$, Eq. 5 gives for $\tau > 0$ the effective residue

$$\tilde{r}_\phi(\tau) = 2\int_{s^*(\tau)}^{1} \phi(s)(1-s)\,\mathrm{d}s\,, \qquad s^*(\tau) = \max\left(0, 1 - \frac{\tau_0}{\tau}\right). \tag{8}$$

We compare two simple quadratic profiles, one enriched near the center $\phi_+$ and the other $\phi_-$ near the wall:

$$\phi_\pm(s) = 1 \pm [\tfrac{1}{2} - s]. \tag{9}$$

They are depicted in Figure 2A. Both profiles range from 0.5 to 1.5 and have a cross-sectional mean of 1, so their measured inlet AIFs are the same. With $q = \min(1, \tau_0/\tau)$ for $\tau > 0$, Eq. 8 gives (writing $\tilde{r}_\pm, \kappa_\pm, R_\pm$ for $\phi_\pm$) $\tilde{r}_+ = \frac{q^2}{2} + \frac{2q^3}{3}$ and $\tilde{r}_- = \frac{3q^2}{2} - \frac{2q^3}{3}$, so that $\kappa_+ = \frac{7}{6}$ and $\kappa_- = \frac{5}{6}$ (Figure 2C). The center-enriched profile $\phi_+$ places more tracer on fast streamlines and therefore delivers more tracer flux. Figure 2D shows that the normalized residues both start at 1 but then separate; at $\tau = 2\tau_0$, for example, $R_+ = \frac{5}{28} \approx 0.18$ and $R_- = \frac{7}{20} = 0.35$.

For a tube with $\tau_0 = 1$ s, we computed tissue curves for both profiles with two kinds of arterial input. The first was a causal gamma-variate bolus, $c_A(t) \propto t^2 e^{-t}$ with $t$ in seconds, representing the contrast bolus of DSC, DCE, or CT perfusion. The second was the rectangular input assumed in the continuous or pseudo-continuous ASL kinetic model (2), labeled at the inlet: its width is the label duration, and multiplying the residue by $e^{-\tau/T_1}$ accounts for intravascular longitudinal relaxation. Because the tissue curve for a rectangular input of duration $D$ integrates the residue function over a sliding window of that width, comparing labels of 0.1, 0.5, and 2 s shows how the profile difference depends on input duration relative to the fastest transit time. ASL curves were computed without and with relaxation ($T_1 = 1.65$ s for blood at 3 T (12)). Curves were evaluated at 0.5-ms intervals, and differences between the center- and wall-enriched tissue curves are reported as percentages of the uniform-inlet ($\phi = 1$) peak under the same relaxation condition.

Despite identical measured inlet AIFs, the two profiles produced different tissue curves (Figure 2B,E). For the gamma-variate input, the curves differed by up to 23.3% of the uniform-inlet peak, and their time to peak by 0.30 s. Because $\kappa_+ \neq \kappa_-$, the amplitude difference combines unequal tracer delivery with the difference in residue shape; the time-to-peak difference reflects shape alone, because scaling a residue by a constant cannot move the peak of the tissue curve, and confirms that the normalized residues also differ (Figure 2D). Both effective residues integrated to $V/F = 2\tau_0$ (2 s; Figure 2C), as required by Eq. 5, so the larger initial value of the center-enriched residue was balanced by its faster decline.

The rectangular inputs show how this difference depends on input duration (Figure 2F,G). For labels shorter than $\tau_0$, no label has left the tube when labeling ends, and the maximum difference reached $\kappa_+ - \kappa_- = 1/3$ of the uniform-inlet peak (33.3%) with or without relaxation, because both tissue curves are then exactly $\kappa_\phi$ times the uniform-inlet curve until label on the fastest streamline reaches the outlet. Without relaxation, longer labels drive the tissue curves toward a common steady state set by the profile-independent area $V/F$ (Eq. 5), so separation peaks on the rising edge (25.0% for a 2-s label). With relaxation, tracer that has spent less time in the tube is weighted more heavily, so the plateaus no longer coincide, and the 2-s separation is 28.0%.

## Discussion

We demonstrated that blood velocity varies across the lumen according to fluid mechanics and consequently a residue cannot be uniquely defined from a cross-sectionally averaged AIF devoid of spatial information about tracer concentration and transit times. For a fixed vascular geometry and velocity field, the same averaged AIF therefore yields different tissue concentration curves for different inlet distributions, even when it is measured at the inlet of the modeled region. The averaged AIF alone cannot distinguish such inlet cross-sectional distributions, which clinical perfusion images do not resolve, causing the ambiguity in the traditional AIF-residue formulation of tracer kinetics.

Consequently, it is difficult to interpret temporal deconvolution results as tissue blood flow, except that the area under the deconvolved response is profile-independent, because $\int_0^\infty \tilde{r}_\phi(\tau)\,\mathrm{d}\tau = V/F$ (Eq. 5). Deconvolving $c_T$ with the area-mean $c_A$ recovers $f \cdot \tilde{r}_\phi = \kappa_\phi \cdot f \cdot R_\phi$, rather than $f$ times a residue function with unit initial value. In the straight-tube example, the peak of this response is $\kappa_\phi \cdot f$. Taken as an estimate of $f$, this peak is biased by ±17%: 7/6 of the true flow for the center-enriched inlet and 5/6 for the wall-enriched inlet, although geometry and flow are identical. In this non-mixing tube, a mean transit time (MTT) computed as the area-based blood volume divided by the peak-based $f$ is scaled by $1/\kappa_\phi$. Referencing the AIF to tracer flux instead of area, as indicator-dilution theory requires (10,13), removes this flow scale factor $\kappa_\phi$ while scaling the area-based blood volume by $1/\kappa_\phi$, but does not remove the shape change of the normalized residue (Eq. 6). When the AIF is measured upstream, the recovered response also absorbs transport between the AIF site and the tissue: a pure delay shifts the response, and dispersion also lowers its peak (14,15), so that flow read from the peak is underestimated.

This ambiguity cannot be removed by AIF delay or dispersion corrections, which address only the intervening artery (16-19). The ambiguity remains even for an AIF measured at the inlet of the modeled region, with no intervening artery to correct for. The effective residue is defined only once an inlet profile is specified (Eq. 5), and the measured AIF does not determine that profile. An analogous dependence is expected whenever a DCE-MRI tissue model is driven by a measured scalar AIF (6).

This work has several limitations. The analysis assumed steady axial (and, in the example, parabolic) flow without radial diffusion; in vivo, pulsatility, vessel curvature, branching, and microvascular mixing set how long radial tracer gradients persist. In the phantom, manual Y-connector injection can also introduce transverse asymmetry. Because the signal was not converted to concentration and view sharing lengthens the time window of each volume, these images show where the tracer is but not how much, keeping this description qualitative. Quantifying two spatially different input distributions with the same AIF will require careful calibration. The effects need to be assessed in future patient studies, such as in evaluating ischemic stroke (20).

## Conclusions

A cross-sectionally averaged AIF does not uniquely determine the residue function, even for a fixed vascular geometry and velocity field and an AIF measured at the inlet of the modeled region. Identical AIFs can produce different tissue concentration curves. The residue function referenced to a measured scalar AIF depends not only on vascular properties but also, in both amplitude and temporal shape, on the inlet tracer distribution.

## Acknowledgments and Funding

This work was supported by the National Institutes of Health (R01EB034755 and R01NS135205).

## Conflict of Interest

PS and YW own equity in Medimagemetric LLC.

## Data Availability Statement

The phantom MRI data and analysis code supporting the findings of this study are available from the corresponding author upon reasonable request.

## Supporting Information

Additional supporting information is provided in the online Supporting Information PDF: Section S1 (Derivations of Eq. 1 and the straight-tube model, Eqs. S1–S3), Section S2 (Reproducibility across all three bolus injections, Figures S1–S2), and Section S3 (Lumen geometry extraction, Figure S3).

## Figures

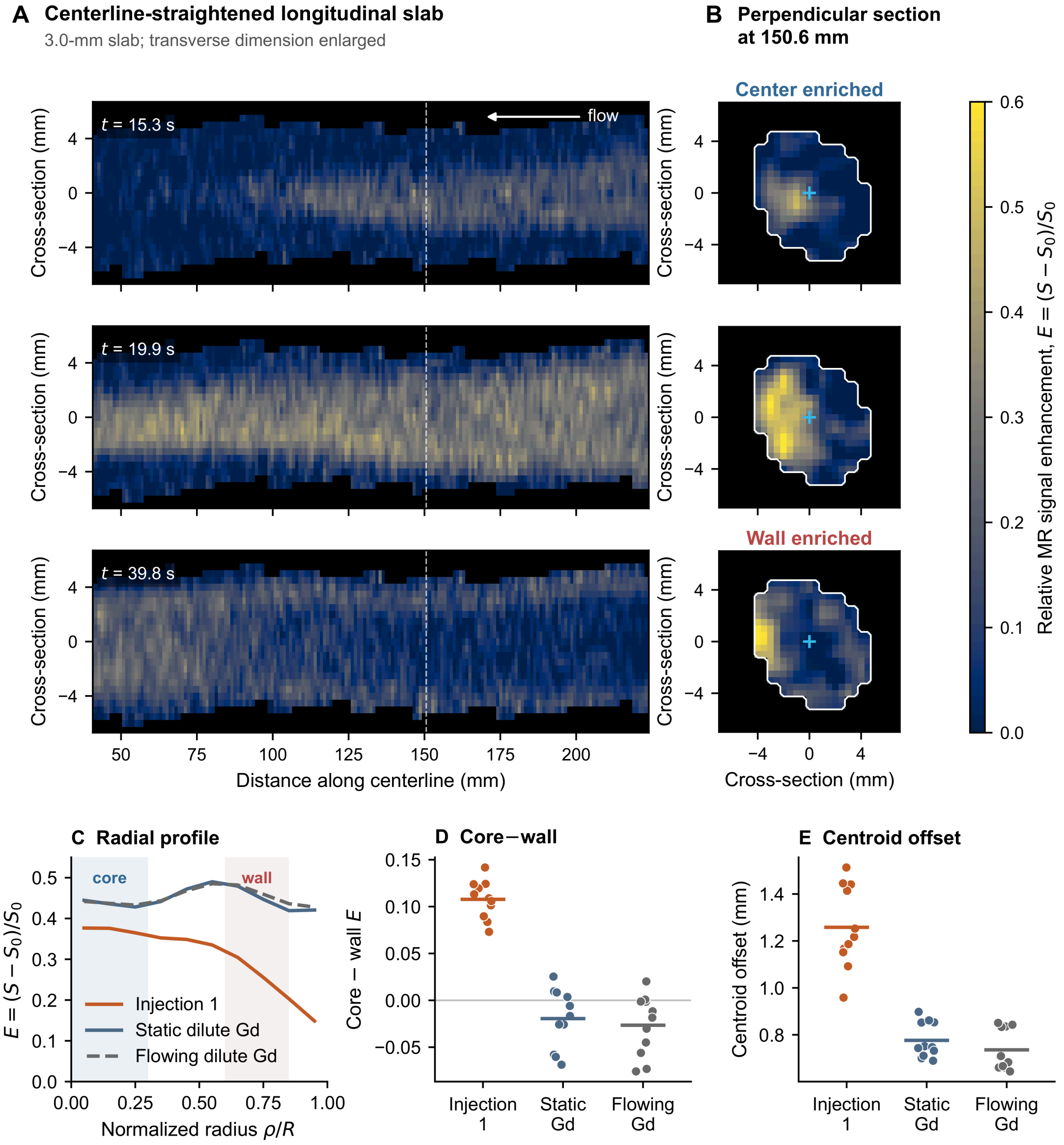

A Centerline-straightened longitudinal slab
3.0-mm slab; transverse dimension enlarged
t = 15.3 s
t = 19.9 s
t = 39.8 s
flow
Cross-section (mm)
Distance along centerline (mm)
B Perpendicular section at 150.6 mm
Center enriched
Wall enriched
Relative MR signal enhancement, $E = (S - S_0)/S_0$
C Radial profile
core
wall
Injection 1
Static dilute Gd
Flowing dilute Gd
$E = (S - S_0)/S_0$
Normalized radius $\rho/R$
D Core−wall
Core − wall $E$
E Centroid offset
Centroid offset (mm)
Injection 1
Static Gd
Flowing Gd

Figure 1. Cross-sectional enhancement during bolus passage and uniform filling. (A) Longitudinal (3.0-mm slab; dashed line marks the 150.6-mm section) and (B) perpendicular bolus maps at 15.3 s (top, center enriched), 19.9 s (middle), and 39.8 s (bottom, wall enriched; largest wall-minus-center difference at this section), relative to the pre-bolus water baseline of the injection acquisition. White contours show the water-derived lumen mask, and the cyan cross marks its center. (C) Eleven-station mean radial profiles (shading marks core, $\rho/R < 0.30$, blue; inner wall, $0.60 \leq \rho/R < 0.85$, red), (D) central-minus-wall ("core–wall") enhancement, and (E) centroid offsets of positive enhancement on the local upslope of Injection 1 and for both uniform gadolinium fills, computed with one common flowing-water baseline and fixed elliptical geometry. Dots show individual stations ($n = 11$) along the tube within the same experiment; horizontal bars show station means.

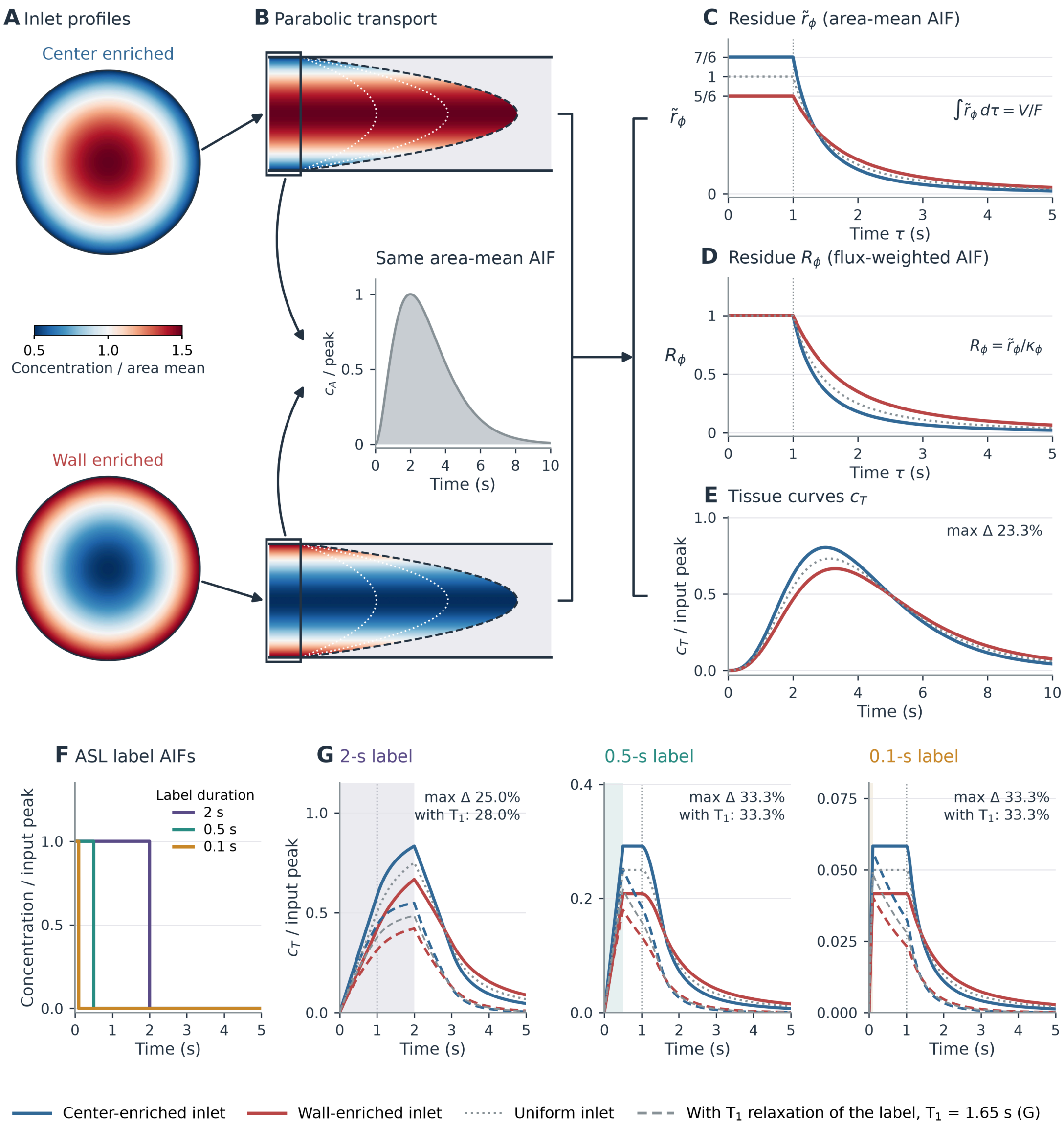

A Inlet profiles
Center enriched
Wall enriched
0.5 1.0 1.5
Concentration / area mean
B Parabolic transport
Same area-mean AIF
$c_A$ / peak
Time (s)
C Residue $\tilde{r}_\phi$ (area-mean AIF)
$\int \tilde{r}_\phi \, d\tau = V/F$
Time $\tau$ (s)
D Residue $R_\phi$ (flux-weighted AIF)
$R_\phi = \tilde{r}_\phi / \kappa_\phi$
Time $\tau$ (s)
E Tissue curves $c_T$
max Δ 23.3%
$c_T$ / input peak
Time (s)
F ASL label AIFs
Label duration
2 s
0.5 s
0.1 s
Concentration / input peak
Time (s)
G 2-s label
max Δ 25.0%
with $T_1$: 28.0%
$c_T$ / input peak
0.5-s label
max Δ 33.3%
with $T_1$: 33.3%
0.1-s label
max Δ 33.3%
with $T_1$: 33.3%
Center-enriched inlet
Wall-enriched inlet
Uniform inlet
With $T_1$ relaxation of the label, $T_1$ = 1.65 s (G)

Figure 2. Identical measured inlet AIFs conceal different residues. (A) Quadratic inlet profiles, enriched near the center or near the wall. (B) In side views of the tube, each profile is carried along axial streamlines of the parabolic flow behind a parabolic bolus front (dashed; white dotted curves show constant-entry-time isochrones), and both give the same area-mean AIF at the inlet (box). (C) Effective residues $\tilde{r}_{\phi}$, referenced to the measured area-mean AIF, start at 7/6 and 5/6 and cross, because both have the same area (Eq. 5). (D) Normalized residues $R_{\phi} = \tilde{r}_{\phi}/\kappa_{\phi}$, referenced to the flux-weighted AIF, all start at 1 but still differ in shape. (E) Resulting tissue curves for the gamma-variate AIF of B, $c_A(t) \propto t^2 e^{-t}$. The maximum absolute difference between tissue curves is 23.3% of the uniform-inlet peak, with a 0.30-s difference in time to peak. (F) Rectangular ASL label AIFs of 2, 0.5, and 0.1 s and (G) the resulting tissue curves without (solid) and with (dashed) longitudinal relaxation of the label ($T_1 = 1.65$ s); shading marks the labeling period, and the dotted vertical line marks $\tau_0$ in C, D, and G. Max Δ is the maximum difference between the center- and wall-enriched tissue curves, as a percentage of the uniform-inlet peak under the same relaxation condition; for labels shorter than $\tau_0$, the difference equals $\kappa_+ - \kappa_- = 1/3$. In C, D, E, and G, gray curves are for a uniform inlet. Concentrations are normalized to the input peak; residues are dimensionless. $\tau_0 = 1$ s.

# Supporting Information

## A scalar arterial input function does not uniquely determine the perfusion residue: Theory and phantom MRI

Yuyang Gan, Dominick Romano, Maneesh John, Renjiu Hu, Qihao Zhang,
Benjamin Weppner, Alexey V. Dimov, Pascal Spincemaille, Yi Wang

Section S1 derives Eq. 1 from tracer mass conservation and gives the steps behind Eqs. 4–6, 8, and 9. Section S2 reports all three bolus injections, and Section S3 summarizes how the lumen geometry of the phantom was extracted. Equation, section, and figure numbers without the prefix S refer to the main text; reference numbers refer to the list below.

## S1. Derivations

### S1.1. Mass balance and Eq. 1

Integrating the tracer transport equation over a vascular or tissue region of volume $V$ (1) gives the mass balance for the tracer mass $m(t) = V\, c_T(t)$:

$$\frac{\mathrm{d}m}{\mathrm{d}t} = J_{\mathrm{in}}(t) - J_{\mathrm{out}}(t). \tag{S1}$$

The standard indicator-dilution model (2) assumes that (i) the entering flux is $J_{\mathrm{in}} = F\, c_A$, making $c_A$ implicitly the flux-weighted inlet concentration, and (ii) transport is linear and time invariant, with a single transit-time density $h(\tau)$ independent of entry position, so that $J_{\mathrm{out}}(t) = F \int_0^t c_A(t')\, h(t - t')\,\mathrm{d}t'$. Integrating Eq. S1 for an initially tracer-free region yields Eq. 1 with $f = F/V$ and $r(\tau) = 1 - \int_0^\tau h(\tau')\,\mathrm{d}\tau'$. When velocity varies across the lumen and $c_A$ is measured as a cross-sectional area mean, neither assumption holds automatically (3).

### S1.2. Tube model

In the straight tube, the true entering flux is $J_{\mathrm{in}} = \int_S u\, c(0,\rho,\theta,t)\,\mathrm{d}S = F\, c_F(t)$, where $S = \pi R^2$ is the cross-sectional area and $c_F$ is the flux-weighted inlet concentration. For a time-invariant inlet profile $c(0,\rho,\theta,t) = \phi(\rho,\theta)\, c_A(t)$ with unit area mean and $c_A(t') = 0$ for $t' < 0$, substituting $c_A(t - z/u) = \int_{-\infty}^{\infty} c_A(t-\tau)\,\delta(\tau - z/u)\,\mathrm{d}\tau$ into Eq. 3 yields Eqs. 4 and 5. For $\tau$ shorter than the fastest transit time, no tracer has exited ($H = 1$ on every streamline), and the initial value of the effective residue is

$$\kappa_\phi = \frac{1}{F}\int_S \phi\, u\,\mathrm{d}S = \frac{c_F}{c_A}, \qquad \kappa_\phi - 1 = \frac{\mathrm{cov}_S(\phi,u)}{\bar{u}}, \tag{S2}$$

where $\bar{u} = F/S$ is the mean velocity and $\mathrm{cov}_S$ the covariance over the cross-section. Thus $\kappa_\phi = 1$ whenever $\phi$ is uniform or uncorrelated with $u$, and $\kappa_\phi > 1$ when tracer is enriched on faster streamlines. Whereas the effective residue always integrates to $V/F$, the normalized residue integrates to the mean transit time of the tracer itself, $\int_0^\infty R_\phi(\tau)\,\mathrm{d}\tau = V/(\kappa_\phi F)$.

For steady parabolic flow, the substitution $s = (\rho/R)^2$ reduces Eq. 5 to Eq. 8, and $w = 1 - s$ gives $\tilde{r}_\phi(\tau) = 2\int_0^q \phi(1-w)\, w\, \mathrm{d}w$ with $q = \min(1, \tau_0/\tau)$ for $\tau > 0$. For a uniform inlet, $\tilde{r}_1(\tau) = q^2$. For the quadratic profiles of Eq. 9, the integral yields $\tilde{r}_\pm$ in the main text and the normalized residues

$$R_+(\tau) = \frac{3q^2 + 4q^3}{7}, \qquad R_-(\tau) = \frac{9q^2 - 4q^3}{5}, \tag{S3}$$

with tracer mean transit times of $12\tau_0/7$ and $12\tau_0/5$, respectively, compared with $2\tau_0$ for a uniform inlet.

## S2. Reproducibility across all three bolus injections

At approximately 200 mL/min in the 9.53-mm tube (Reynolds number Re ≈ 450, well within the laminar regime), the injection acquisition included three manual boluses spaced approximately 60 s apart, and all three reproduced the same spatial evolution: central enhancement dominated during bolus arrival, followed by near-wall lingering during washout (Figure S1A,B). Before bolus 1, the small lumen-mean offset vs. flowing water (0.018; core minus wall 0.006) reflects a slight inter-scan shift (Figure S2A–C). Because washout was incomplete between consecutive injections, residual gadolinium lingered near the wall before the second and third boluses (Figure S1C, top row; Figure S2A–C). When referenced to the common flowing-water baseline, this wall-weighted residual partially offset the upslope central-minus-wall difference of the later boluses (0.076 and 0.063 vs. 0.108 for the first bolus; Figure S1C, middle row). When referenced instead to each bolus's own pre-bolus frames to subtract the preceding residual, all three boluses yielded nearly identical central-minus-wall differences (0.099, 0.100, and 0.105; Figure S1C, bottom row). Under either baseline, central-minus-wall enhancement was positive at all 11 stations for every bolus (Figure S2E,H).

## S3. Lumen geometry extraction

Because the tube was inclined by 3.3° to purge air bubbles (rising 15 mm over the 259-mm field of view), native slices cut its lumen obliquely, and the lumen left the 20-mm imaging slab near one end (Figure S3A,B). To avoid bias from the nonuniform enhancement of the bolus itself, the tube centerline and a consensus elliptical lumen boundary (median semi-axes 5.16 and 3.98 mm) were fitted solely from the static- and flowing-water acquisitions, on perpendicular cross-sections resampled on a 0.4-mm grid (Figure S3C,D). Of 20 stations spaced 8.1 mm apart along the tube, the 11 fully contained within the slab were analyzed and the 9 clipped stations were excluded (Figure S3A). Radial metrics used the normalized elliptical radius $\rho/R$ within the fixed consensus boundary (Figure S3E).

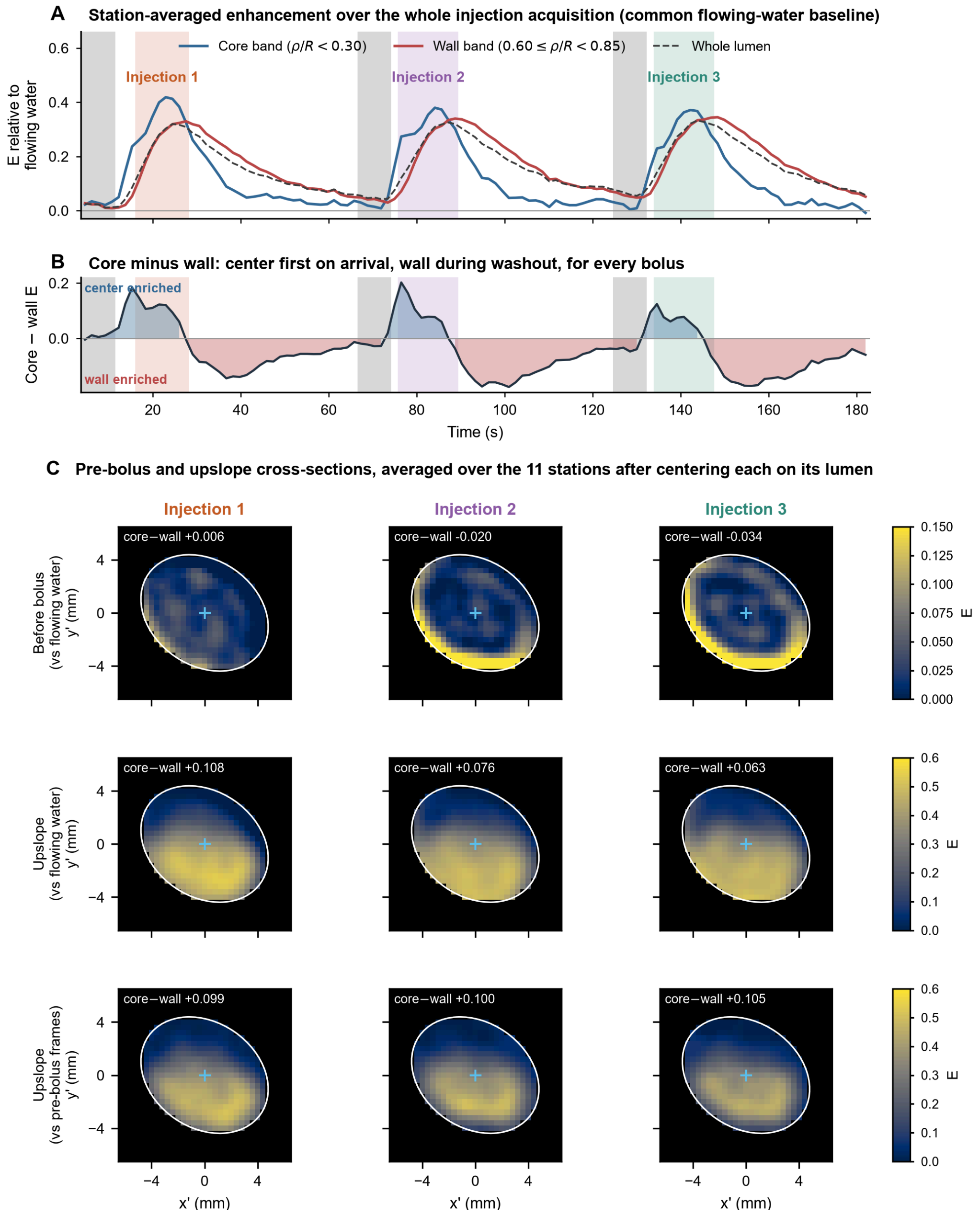

Figure S1. All three boluses. (A) Central-band (blue), wall-band (red) and whole-lumen (dashed) enhancement over frames 4–120, averaged over the 11 stations and referenced to flowing water; gray, pre-bolus frames; colored, span of station-wise upslope frames across the 11 stations. (B) Central minus wall. (C) Cross-sections averaged over the stations after centering each on its lumen: residual before each bolus (top; note the color scale), and the upslope referenced to flowing water (middle; the main-text reference) or to the bolus's own pre-bolus frames (bottom); $x'$ and $y'$ are coordinates in the perpendicular plane. Numbers give the central-minus-wall difference.

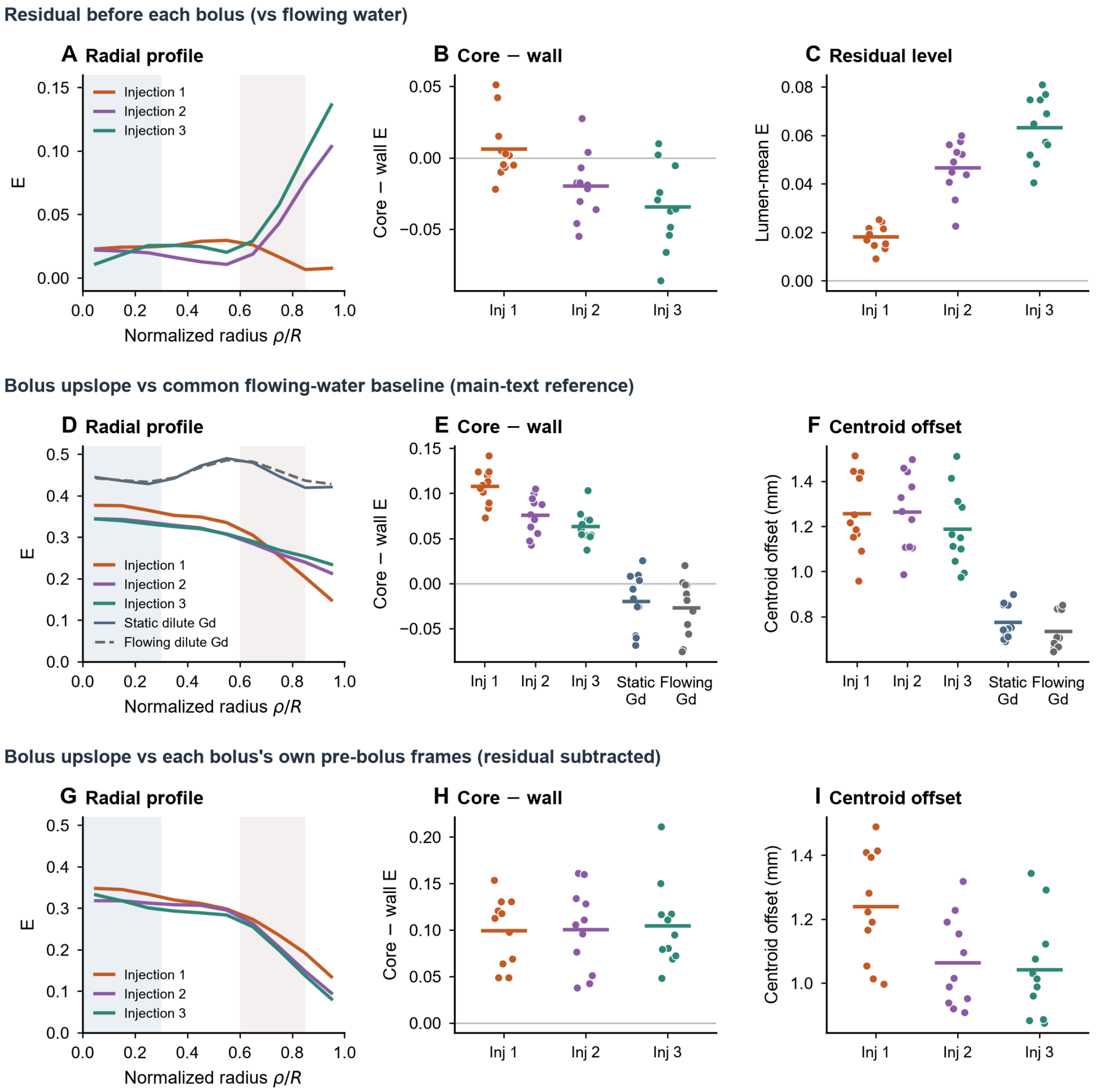

Figure S2. Station metrics for the residual before each bolus (A–C) and for the upslope referenced to flowing water (D–F, with the uniform fills; bolus 1 reproduces Figure 1C–E) or to each bolus's own pre-bolus frames (G–I). Dots, stations; bars, station means.

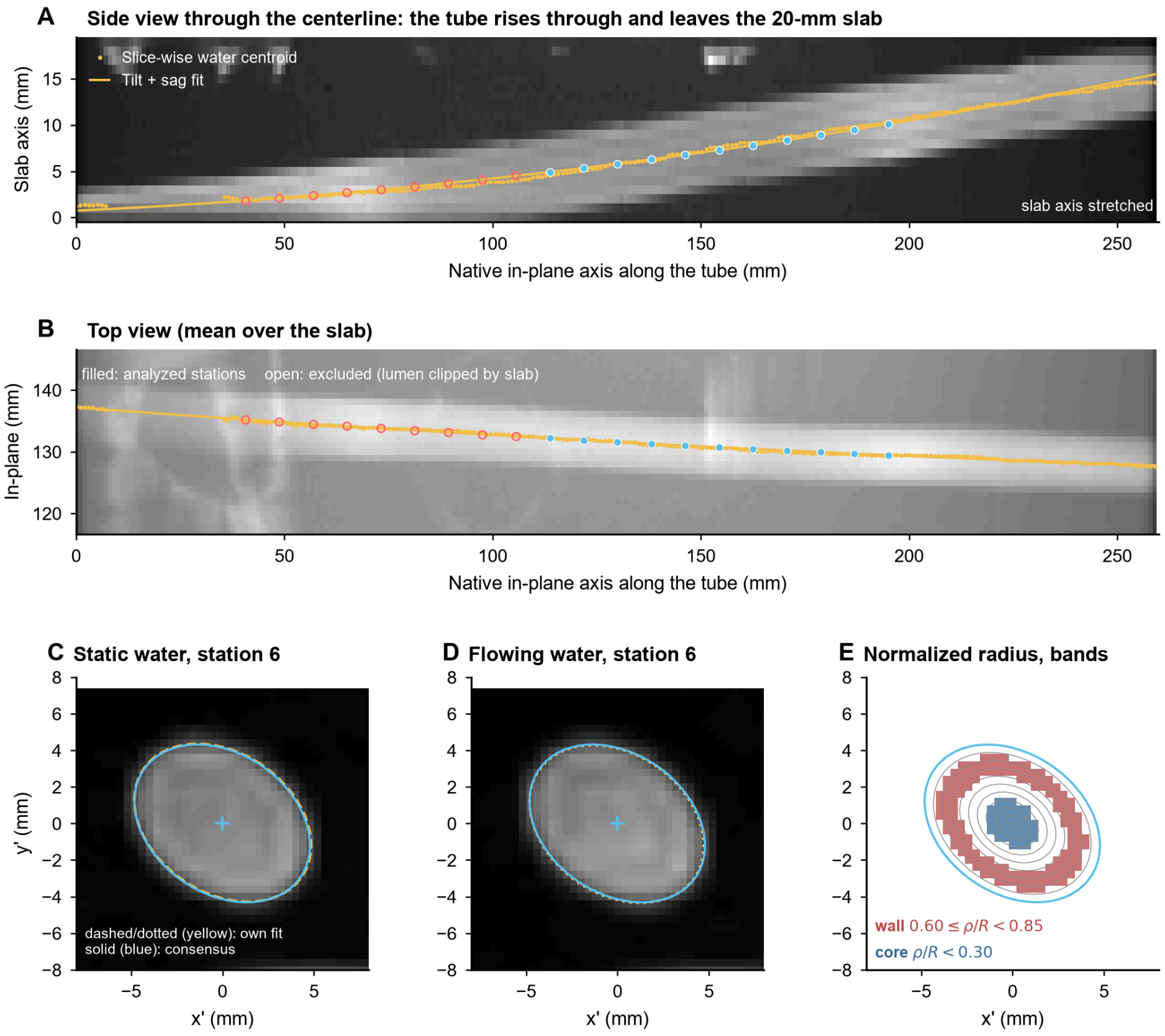

Figure S3. Lumen geometry. (A) Side view of the combined water images through the centerline, with the slab axis stretched: the tube rises through the 20-mm slab and leaves it at the low end. Dots, slice-wise lumen centroids; line, tilt-plus-sag fit. (B) Top view (mean over the slab); in (A, B), filled blue and open red circles mark the 11 analyzed and 9 clipped stations. (C, D) Static- and flowing-water images in the perpendicular plane at the sixth analyzed station, with each image's own ellipse fit (dashed/dotted yellow) and the consensus ellipse (solid blue); $x'$ and $y'$ are coordinates in that plane. (E) Normalized radius $\rho/R$, with the central ("core", blue) and inner wall (red) bands.